\documentstyle[12pt]{article}
\begin{document}
\vskip 2cm
\begin{center}
{\large {\sf  COMMENTS ON THE DUAL-BRST SYMMETRY}}

\vskip 3.0cm

{\sf S. Krishna$^{(a)}$, A. Shukla$^{(a)}$, R. P. Malik$^{(a,b)}$}\\
$^{(a)}$ {\it Physics Department, Centre of Advanced Studies,}\\
{\it Banaras Hindu University, Varanasi - 221 005, (U.P.), India}\\

\vskip 0.1cm

{\bf and}\\

\vskip 0.1cm

$^{(b)}$ {\it DST Centre for Interdisciplinary Mathematical Sciences,}\\
{\it Faculty of Science, Banaras Hindu University, Varanasi - 221 005, India}\\
{\small {\sf {e-mails: skrishna.bhu@gmail.com; ashukla038@gmail.com; malik@bhu.ac.in}}}

\end{center}

\vskip 2cm

\noindent
{\bf Abstract:} 
In view of a raging controversy on the topic of dual-Becchi-Rouet-Stora-Tyutin (dual-BRST/co-BRST) and anti-co-BRST symmetry transformations in the context of four (3+1)-dimensional (4D) Abelian 2-form and 2D (non-)Abelian 1-form gauge theories, we attempt, in our present short note, to settle the dust by taking the help of mathematics of differential geometry, connected with the Hodge theory, which was the original motivation for the nomenclature
of ``dual-BRST symmetry'' in our earlier set of works. It has been claimed, in a recent set of papers, that the co-BRST symmetries are {\it not} independent of the BRST symmetries. We show that the BRST and co-BRST symmetries are independent symmetries in the same fashion as the exterior and co-exterior derivatives are independent entities belonging to the set of de Rham cohomological  operators of differential geometry.\\



\vskip 2cm


\newpage
The local symmetries, generated by the first-class constraints in the language of 
Dirac's prescription for the classification scheme [1,2], are the key signatures 
of a gauge theory. The above symmetries are traded with the Becchi-Rouet-Stora-Tyutin 
(BRST) and anti-BRST symmetries when one performs 
the covariant canonical quantization of the gauge theories. The physicality condition
$Q_b|phys\rangle=0$, in terms of the nilpotent $(Q_b^{2}=0)$ and conserved $(\dot Q_b=0)$ BRST charge $Q_b$,
leads to the annihilation of the physical states by the operator form of the first-class constraints (belonging to the original gauge theory). This observation turns out to be consistent with the Dirac's prescription for the quantization of physical systems with constraints as it fulfills the key quantization requirements [1,2].

The nilpotency property $(Q_b^{2}=0)$ and physicality criteria $Q_b|phys \rangle=0$
are the two essential ingredients that lead to the physical realization of exteriar derivative $d$
(with $d = dx^{\mu}\partial_\mu, d^{2}=0$) of the de Rham cohomological operator of differential geometry [3-6] in the language of BRST charge $Q_b$. In a set of papers [7-10], we have provided the physical realizations of the other {\it two} de Rham cohomological operators of differential geometry in the context of 4D Abelian 2-form gauge theory.
The latter two cohomological operators are the co-exterior derivative $ \delta = -*d*$ 
(with $\delta^{2}=0 $) and the Laplacian operator $\bigtriangleup = (d + \delta)^{2}= d\delta + \delta d$
where $*$ is the Hodge duality operation on the 4D flat Minkowaski spacetime manifold.
We have also provided the physical meaning of $\delta= -*d*$ and $\bigtriangleup$ for the 
2D (non-)Abelian 1-form free gauge theory (without any interaction with matter fields)
[11-13] and interacting Abelian $U(1)$ gauge theory (QED) with the Dirac fields [14,15]

In a recent set of papers on the 4D Abelian 2-form gauge theory [16,17]and earlier papers 
on 4D as well as 2D Abelian 1-form gauge theories [18-20], the authors have claimed that the dual-BRST symmetry is {\it not} an independent symmetry transformation.
Rather, this symmetry can be obtained from the BRST symmetry transformation by performing a canonical transformation in the phase space of the ghost sector. The purpose of our present short note is 
to settle the dust that has been generated by the above cited papers as far as the linear (in)dependence of the BRST and co-BRST symmetries is concerned. We show, furthermore,
that the existence of the dual-BRST symmetry and a bosonic symmetry (corresponding to the Laplacian operator) are on the firm foundations of the mathematics of differential geometry
and they have their own independent identity.

Let us begin with the following coupled (but equivalent) Lagrangian densities [10]
\begin{eqnarray}
{\cal L}_{(B, {\cal B})} &=& \frac{1}{2} {\cal B}_\mu  {\cal B}^{\mu}- {\cal B}^{\mu} \Bigl (\frac{1}{2}\varepsilon_{\mu\nu\eta\kappa} \partial^{\nu} B^{\eta\kappa} + \frac{1}{2}  \partial_\mu \phi_2  \Bigr )  
+ B^{\mu} \Bigl (\partial^{\nu} B_{\nu\mu} + \frac{1}{2}  \partial_\mu \phi_1  \Bigr ) - \frac{1}{2} B_\mu  B^{\mu} \nonumber\\
&+& \partial_\mu \bar \beta \partial^{\mu} \beta + 
(\partial_\mu \bar C_\nu - \partial_\nu \bar C_\mu ) (\partial^{\mu} C^{\nu})
+ (\partial \cdot C - \lambda ) \rho + (\partial \cdot \bar C + \rho ) \lambda ,
\end{eqnarray}
\begin{eqnarray}
{\cal L}_{(\bar B, {\bar {\cal B}})} &=& \frac{1}{2} {\bar {\cal B}}_\mu  {\bar {\cal B}}^{\mu} - {\bar {\cal B}}^{\mu} \Bigl (\frac{1}{2}\varepsilon_{\mu\nu\eta\kappa} \partial^{\nu} B^{\eta\kappa} - \frac{1}{2}  \partial_\mu \phi_2  \Bigr ) + \bar B^{\mu} \Bigl (\partial^{\nu} B_{\nu\mu} - \frac{1}{2} \partial_\mu \phi_1  \Bigr ) - \frac{1}{2} {\bar B}_{\mu} {\bar B}^{\mu} \nonumber\\ &+& \partial_\mu \bar \beta \partial^{\mu} \beta + 
(\partial_\mu \bar C_\nu - \partial_\nu \bar C_\mu ) (\partial^{\mu} C^{\nu})
+ (\partial \cdot C - \lambda ) \rho + (\partial \cdot \bar C + \rho ) \lambda ,
\end{eqnarray}
where the 2-form $[B^{(2)} = \frac {1}{2!}\;(dx^\mu \wedge dx^\nu) B_{\mu\nu}]$
defines the antisymmetric tensor gauge field $B_{\mu\nu}$, the Lorentz vector fermionic
($C_\mu^{2}=0$, $\bar C_\mu^{2}=0,\; C_\mu \bar C_\nu + \bar C_\nu C_\mu=0$, etc.)
fields ($\bar C_\mu$)$ C_\mu$ are the (anti-)ghost fields, the bosonic fields 
($\bar \beta$)$\beta$ are the ghost-for-ghost (anti-)ghost fields and the auxiliary 
(anti-)ghost fields $(\rho)$ $\lambda $ are also needed for the unitarity in the theory.
The Lorentz vector fields $B_\mu$ and $\cal B_\mu$ are the Nakanishi-Lautrup type auxiliary fields which are required for the linearization of the kinetic and 
gauge-fixing terms for the 2-form gauge field $B_{\mu\nu}$. 
The massless ($\Box \phi_1 = 0$, $\Box \phi_2 = 0$) scalar fields $\phi_1$ and $\phi_2$
are also present in the 4D Abelian 2-form gauge theory (within the framework of BRST formalism).

It has been shown (see, e.g. [10]) that the above coupled Lagrangian densities respect
{\it six} continuous symmetry transformations. Under the (anti-)BRST symmetry transformations
$s_{(a)b}$, the kinetic term (owing its origin to the exterior derivative $d$) remains invariant. On the other hand, it is the gauge-fixing term (owing its origin to the co-exterior derivative $\delta = - *d*$) that remains invariant under the (anti-)co-BRST symmetry transformations $s_{(a)d}$. The analogue of the Laplacian operator
($\bigtriangleup = \{ d,\delta \}$) is a bosonic symmetry
$s_w = \{s_b, s_d \} = - \{ s_{ab}, s_{ad} \}$, under which, the ghost terms remain invariant.The theory also respects the ghost symmetry $s_g$, under which, only the (anti-)ghost fields transform.

The above cited continuous symmetry transformations, in their operator form, obey the following interesting algebra (see, e.g. [10]):  
\begin{eqnarray}
s_{(a)b}^{2}= 0, \quad s_{(a)d}^{2}= 0, \quad \{ s_b, s_d \} = s_w = -\{ s_{ab}, s_{cd} \}, \nonumber\\
\bigl[ s_w, s_r \bigr ] = 0, \quad r= b, ab, d, ad, g, w, \quad
i\;\bigl [s_g, s_b \bigr ] = + s_b, \nonumber\\ i \;\bigl [s_g, s_{ad}\bigr ]= +s_{ad}, 
\quad i \;\bigl [s_g, s_{ab} \bigr ] = - s_{ab}, \quad i \;\bigl [s_g, s_{d} \bigr ] = -s_d,
\end{eqnarray}
which are reminiscent  of the algebra obeyed by the de Rham cohomological operators
\begin{eqnarray}
d^{2} = \delta^{2}=0, \quad \bigtriangleup = (d + \delta )^{2} = \{ d,\delta \} , \nonumber\\
\bigl [\bigtriangleup , \delta  \bigr ] \; =\;  \bigl [\bigtriangleup , d \bigr ] = 0, \qquad \delta \;=\;\pm *d*,
\end{eqnarray}
where $*$ is the Hodge duality operation and the Laplacian operator is the casimir operator. A close look at 
equations (3)
and (4) establishes the fact that there is two-to-one mapping $\Bigl [ i.e.\;(s_b, s_{ad}) \rightarrow d, \; (s_{ab}, s_d)\rightarrow \delta$ and $ \{s_b, s_d \} = - \{ s_{ab}, s_{ad}\}\rightarrow \bigtriangleup \Bigr ]$ between the symmetry transformations and the de Rham cohomological operators of differential geometry.

It is very interesting to point out that the analogue of the Hodge duality $*$ operator [10] is the presence of the following discrete symmetry transformations, namely;
\begin{eqnarray}
B_{\mu\nu} \rightarrow \mp  \frac {i}{2} \; \varepsilon_{\mu\nu\eta\kappa} B^{\eta\kappa}, \quad 
C_\mu \rightarrow \pm i \;\bar C_\mu , \quad \bar C_\mu \rightarrow \pm i \; C_\mu, \nonumber\\
\beta \rightarrow \mp i \; \bar \beta , \quad \bar \beta \rightarrow \pm i \; \beta ,
\quad \phi_1 \rightarrow \pm i \;\phi_2, \quad \phi_2 \rightarrow \mp i \;\phi_1, \nonumber\\
\rho \rightarrow \pm i\; \lambda, \quad \lambda \rightarrow \pm i\; \rho, \quad B_\mu 
\rightarrow \pm i\;{\cal B}_\mu, \quad {\cal B}_\mu \rightarrow  \mp i\; B_\mu,  
\end{eqnarray}
in the theory, under which, the Lagrangian densities (1) and (2)
remain invariant. The importance of the above duality transformations becomes quite transparent in the following relationship
\begin{eqnarray}
s_{(a)d} \; \Psi = \pm *s_{(a)b}* \Psi, \qquad \Psi = 
{\cal B}_{\mu\nu}, B_{\mu\nu}, C_\mu, \bar C_\mu, \beta,\bar\beta, \rho, \lambda, \phi_1, \phi_2, B_\mu,
\end{eqnarray}
where $\Psi$ is the generic field of the theory.
If we take the following off-shell nilpotent BRST and co-BRST transformations [10]
\begin{eqnarray}
s_b B_{\mu\nu} &=& - (\partial_\mu C_\nu - \partial_\nu C_\mu), \;\quad s_b C_\mu = - \partial_\mu \beta,  \;\quad s_b {\bar C}_\mu = - B_\mu,\nonumber\\ s_b \phi_1 &=& - 2\; \lambda , \;\qquad s_b \bar\beta = - \rho,
\;\qquad s_b \bigl [\rho , \lambda, \beta, \phi_2, B_\mu, {\cal B}_\mu \bigr ] = 0,
\end{eqnarray}
\begin{eqnarray}
s_d B_{\mu\nu} &=& - \varepsilon_{\mu\nu\eta\kappa} \partial^{\eta} \bar C^{\kappa}, \;\quad s_d C_\mu = - {\cal B}_\mu, \;\quad s_d {\bar C}_\mu = - \partial_\mu \bar \beta,\nonumber\\ s_d \phi_2 &=&  2\;\rho, \;\quad s_d \beta = - \lambda,
\;\quad s_d \bigl [\rho , \lambda, \bar \beta, \phi_1, B_\mu, {\cal B}_\mu, \partial^{\nu} B_{\nu\mu} \bigr ] = 0,
\end{eqnarray}
it can be explicitly checked that the relationship ($ s_d \Psi = \pm *s_b*\Psi$) is satisfied when the equations (5), (7) and (8) are used properly.

It will be mentioned that the discrete symmetry transformations (5) are the analogue of the $(*)$ operator in the equation (6). Furthermore, it is interesting to point out that 
($\pm $) signs in equation (6) are governed by the two successive operations of the discrete symmetry transformation on any generic field $\Psi$ (see, e.g. [21] for details):
\begin{eqnarray}
*\;(* \; \Psi) = \pm \Psi, \; \qquad \Psi = B_{\mu\nu}, C_\mu, \bar C_\mu, \beta,\bar\beta
, \rho, \lambda, \phi_1, \phi_2, B_\mu, {\cal B}_\mu.
\end{eqnarray}
Thus, we note that the duality relationship between the co-exterior derivative ($\delta$)
and exterior derivative ($d$) (i.e. $\delta = \pm *d*$) is captured by the relationship 
$s_d = \pm *s_b*$. It will be noted that the other relationship $s_{ad} = \pm *s_{ab}*$
is also true when applied on any arbitrary generic field $\Psi$ of the theory. Here
$s_{ab}$ and $s_{ad}$ are the nilpotent anti-BRST and anti-co-BRST symmetry transformation that are quoted in [10].

Let us now focus on the {\it free} 2D Abelian gauge theory (for the sake of simplicity). The Lagrangian density, that respects (anti-)BRST symmetry transformations, is [11-13]
\begin{eqnarray}
{\cal L}_b = {\cal B} E - \frac {{\cal B}^{2}}{2} + B(\partial \cdot A) + \frac {B^{2}}{2} - i\; \partial_\mu \bar C \;\partial^{\mu}C,
\end{eqnarray}
where $B$ and ${\cal B}$ are the (Lorentz scalar) Nakanishi-Lautrup type auxiliary field, $E = \partial_0 A_1 - \partial_1 A_0 = F_{01}$ is the electric field and fermionic $(C^{2}= \bar C^{2}=0,\;C \bar C + \bar C C=0)$
(anti-)ghost fields $(\bar C)C$ are required for the unitarity in the theory. It can be checked that the following nilpotent BRST and co-BRST transformations
\begin{eqnarray}
s_b A_\mu &=& \partial_\mu C, \qquad s_b C = 0, \qquad s_b \bar C = i\;B, \quad s_b {\cal B}= 0,\qquad s_b E = 0, \quad s_b B=0, \nonumber\\
s_d A_\mu &=& -\varepsilon_{\mu\nu}\partial^\nu \bar C, \quad s_d \bar C = 0, \quad s_d  C = i\;{\cal B}, \quad s_d {\cal B}= 0, \quad s_d B = 0, \; s_d (\partial \cdot A) = 0,
\end{eqnarray}
are the symmetry transformations for the Lagrangian density ${\cal L}_b$.

There are a couple of remarks in order. First, it can be checked that the kinetic term $(\frac {1}{2} E^{2} = \frac {1}{2} F_{01}\;F^{01}\equiv {\cal B} E - {\cal B}^{2}/{2})$,
owing its origin to the exterior derivative $d = dx^{\mu} \partial_\mu$, remains invariant under the BRST
symmetry transformation $(s_b)$. On the other hand, it is the 
gauge-fixing term $\Bigl (- \frac {1}{2} (\partial \cdot A)^{2} \equiv B (\partial \cdot A) + B^{2}/{2} \Bigr)$, owing its origin to the co-exterior derivative $(\delta = - *d*)$, that remains invariant under the co-BRST symmetry transformations $(s_d)$. Second, it is interesting to note that, for the generic field $\Phi$, we have
\begin{eqnarray}
s_d \Phi = \pm *s_b*\;\Phi,  \qquad \Phi = E, B, {\cal B}, C, \bar C, A_\mu.
\end{eqnarray}
In the above, the $(*)$ operation is nothing but the following discrete transformations
\begin{eqnarray}
A_\mu &\rightarrow & \mp i\; \varepsilon_{\mu\nu} A^{\nu}, \quad C \rightarrow \pm i\; \bar C,
\quad \bar C\rightarrow \pm i\; C, \quad B \rightarrow \mp  i\; {\cal B}, \nonumber\\
{\cal B}&\rightarrow & \mp  i\;B, \;\qquad (\partial \cdot A)\rightarrow \pm i\;E, 
\;\qquad E\rightarrow \pm  i\;(\partial \cdot A),
\end{eqnarray}
under which, the Lagrangian density ${\cal L}_b$ remains invariant. As discussed earlier,
the $(\pm )$ signs in $s_d \Phi = \pm *s_b*\Phi$ are dictated by the {\it two} successive  operations of the transformations (13) on any generic field $\Phi$, namely;
\begin{eqnarray}
*\;(*\;\Phi) = \pm\;\Phi, \qquad \Phi = C, \bar C, A_\mu, B, {\cal B}, (\partial \cdot A), E.
\end{eqnarray}
It is worth pointing out that, in our earlier works [11-13], the anti-BRST and anti-co-BRST transformations have also been derived and discussed. They also satisfy $s_{ad} = \pm *s_{ab}*\Phi$.

The above discussions provide enough proofs for the 4D Abelian 2-form as well as 2D Abelian 1-form gauge theories (within the framework of BRST formalism) to be the field theoretic models for the Hodge theory. This assertion can be corroborated by the observations that the conserved charges ($Q_r,\; r = w, b, d, ab, ad, g$) of both the theories satisfy the following beautiful relationships [7-16], namely;
\begin{eqnarray}
i\;Q_g\;Q_b |\Psi \rangle_n &=& (n+1)\;Q_b |\Psi \rangle_n, 
\quad i\;Q_g\;Q_{ad} |\Psi \rangle_n = (n+1)\;Q_{ad} |\Psi \rangle_n,\nonumber\\
i\;Q_g\;Q_{ab} |\Psi \rangle_n &=& (n-1)\;Q_{ab} |\Psi \rangle_n,
\quad i\;Q_g\;Q_d |\Psi \rangle_n = (n-1)\;Q_d |\Psi \rangle_n, \nonumber\\
i\;Q_g\;Q_w |\Psi \rangle_n &=& n\; Q_w |\Psi \rangle_n, \;\;\qquad 
i\;Q_g|\Psi \rangle_n = n\;|\Psi \rangle_n.
\end{eqnarray}
The ghost number `$n$' of a state $|\Psi \rangle_n$ is defined through the ghost charge 
$i\;Q_g |\Psi\rangle_n = n\;|\Psi \rangle_n$. It can be noted that
$ Q_b |\Psi\rangle_n, Q_d |\Psi\rangle_n $ and $Q_w |\Psi\rangle_n$ have the ghost numbers equal to $(n+1)$, $(n-1)$
and $n$, respectively. Similarly, the ghost numbers of $ Q_{ad} |\Psi\rangle_n, Q_{ab} |\Psi\rangle_n $ and $Q_w |\Psi\rangle_n$ are also $(n+1)$, $(n-1)$ and $n$, respectively.

The above observations (15) are equivalent to the statement that the degree of a given 
$n$-form $(f_n)$ increases by one when the exterior derivative applies on it (i.e. $d f_n \sim f_{n+1}$). On the contrary, it is well-known
that the co-exterior derivative ($ \delta = - *d*$) decreases the degree of 
a $n$-form ($f_n$) by one when it operates on it (i.e. $\delta f_n \sim f_{n-1}$). It is also well-established fact that the Laplacian operator does not change the degree of a form on which it operates 
(i.e. $\bigtriangleup  f_n \sim f_n$) because of its structure $\Delta = d \delta + \delta d$.

We would like to lay emphasis on the fact that the canonical transformation in the phase space of merely the ghost sector (as claimed in the recent and old papers [16-20]), do not capture all the issues that have been discussed and explained in our works [7-15]. It is self-evident that $\delta$ is expressed in terms of $d$ (i.e. $ \delta = - *d*$).
Thus, we would like to comment that, since $d$ and $\delta = -*d*$ are related, the canonical transformations address only one aspect of this relationship. However, the rest of the properties, associated with $d$ and $\delta = - *d*$, are {\it not} addressed by the above authors in their works [16-20].

Finally, we would like to state that, in the purview of our investigations, the dual-BRST symmetry
exists for any Abelian $p$-form gauge theory where the following relationship:
\begin{eqnarray}
*\;(d\; A^{(p)}) = \delta \;A^{(p)},
\end{eqnarray}
is satisfied. Here $A^{(p)}$ is the Abelian $p$-form gauge field. This is precisely the reason that, for 2D Abelian 1-form and 4D Abelian 2-form gauge theories, we have dual-BRST symmetries. We predict that 6D Abelian 3-form gauge theory would also respect the existence of dual-BRST symmetry (within the framework of BRST formalism) as the relationship (16) is satisfied for it. It is gratifying to state that we have already made some progress 
in this direction [22].
We make a general statement that any arbitrary $p$-form gauge theory would be endowed with the dual-BRST symmetry if the dimensionality (D) of spacetime is $D= 2 p$.

\end{document}